\documentclass[letterpaper]{article}
\usepackage{aaai}
\usepackage{times}
\usepackage{helvet}
\usepackage{courier}
\usepackage{graphicx}
\usepackage{multirow}
\usepackage{subfig}
\usepackage{blindtext}
\usepackage{verbatim}
\usepackage{mathtools}
\newcommand{\ra}[1]{\renewcommand{\arraystretch}{#1}}
\begin{document}
%
\title{FitChat: Conversational Artificial Intelligence Interventions for \\Encouraging Physical Activity in Older Adults}

\author{Nirmalie Wiratunga\textsuperscript{\rm 1}, 
Kay Cooper\textsuperscript{\rm 2}, 
Anjana Wijekoon\textsuperscript{\rm 1}, Chamath Palihawadana\textsuperscript{\rm 1}\\
{\bf \Large Vanessa Mendham\textsuperscript{\rm 2}, Ehud Reiter\textsuperscript{\rm 3}, Kyle Martin\textsuperscript{\rm 1} }\\
\textsuperscript{\rm 1}School of Computing and Digital Media, Robert Gordon University, Aberdeen\\
\textsuperscript{\rm 2}School of Health Sciences, Robert Gordon University, Aberdeen\\
\textsuperscript{\rm 3}Department of Computing Science, University of Aberdeen, Aberdeen\\
}

\maketitle
\begin{abstract}
\begin{quote}
Delivery of digital behaviour change interventions which encourage physical activity has been tried in many forms. Most often interventions are delivered as text notifications, but these do not promote interaction. Advances in conversational AI have improved natural language understanding and generation, allowing AI chatbots to provide an engaging experience with the user. For this reason, chatbots have recently been seen in healthcare delivering digital interventions through free text or choice selection. In this work, we explore the use of voice based AI chatbots as a novel mode of intervention delivery, specifically targeting older adults to encourage physical activity. We co-created ``FitChat'', an AI chatbot, with older adults and we evaluate the first prototype using Think Aloud Sessions. Our thematic evaluation suggests that older adults prefer voice based chat over text notifications or free text entry and that voice is a powerful mode for encouraging motivation.
\end{quote}
\end{abstract}

\section{Introduction}
\noindent 

Presently, the most common method of delivering Digital Behaviour Change Interventions (DBCIs) is via text-based notifications on mobile phones. Despite the popularity of this approach, there is little evidence to indicate that text notifications are effective at promoting positive behaviour change, particularly in the long-term. The main problem is that text notifications offer only one-way communication from the device to the user, meaning explicit interaction is not required. Accordingly, text notifications are easily ignored; fewer than 30\% of received notifications are typically viewed by users with average delays of close to 3 hours~\cite{morrison2018}. There is clearly a need for an alternative approach.

As a communication medium, conversation appeals to all age groups, but arguably more so towards older adults. This group can have difficulties with new technologies and may be more inclined to appreciate the natural interaction offered by conversational dialogue. With this in mind, we posit that conversation (more specifically, voice-based conversation) presents an opportunity to deliver behaviour change interventions to motivate higher levels of adoption and adherence in older adults when compared with traditional approaches. 

Studies have identified the positive effects of text or choice based conversational agents in specific healthcare domains such as weight loss~\cite{stein2017fully,addo2013toward}, alcoholism treatment~\cite{lisetti2011toward,lisetti2013can} and management of mental health conditions~\cite{morris2018towards,inkster2018empathy,suganuma2018embodied}. 
However, there remain few conversational applications which target general fitness.
With this in mind, we plan to exploit the advances in conversational AI and explore conversation as a form of delivering interventions in general fitness applications, specifically for older adults. In recent years, voice-based conversation assistants have been promoted through easy consumer access in smart home devices and smart phones (e.g. Alexa, OK Google). 
This means that our work is well-placed to investigate conversation as an alternative to current text-based intervention methods.

Our vision is to develop an ubiquitous and proactive system that delivers behaviour change interventions in the form of conversation aimed at promoting physical activities in older adults. 
We start by bringing together end-users from the community through co-creation workshops to help understand what are meaningful conversational interventions and therein develop and design a prototype.
In this paper we present findings from three workshops that shaped the development of a smart phone application integrating a voice based conversational agent. 
A key contribution involves the personalisation of conversation on the basis of a user's context which can be formed using both explicit (e.g. user-entered information) and implicit (e.g. activity data from mobile / wearable devices) information. 
Related to this is the opportunity to use conversational AI to recognise barriers and respond with appropriate motivational dialogue. 
We evaluate the first phase of the intervention with \emph{Think Aloud} sessions aimed at seeking answers to the following questions:
\begin{itemize}
\item What conversational skills are most effective in a fitness chatbot?
\item Can a voice based conversational intervention motivate positive behaviour change?
\item Can we create an experience that ensures full engagement to motivate long-term adherence?
\end{itemize} 

This paper is organised as follows. The Co-creation section presents details of the co-creation workshops and how they contributed to the incremental design of the conversational agent. Afterwards we present the System Architecture, followed by a section discussing our Think Aloud Sessions and presenting a thematic analysis of their outcomes. Related literature in conversational AI is presented before our Conclusions.

\begin{table*}[ht]
    \caption{Intent Templates and Slots}
    \label{tbl:temp}
    \centering
    \ra{1.2}
    \begin{tabular}{@{\hspace{5pt}} l @{\hspace{15pt}} l @{\hspace{15pt}} l @{\hspace{15pt}} l @{\hspace{5pt}}}
    \hline
    Intent & Task & Data Extraction Slots & Contextualisation Slots\\
    \hline
    \multirow{2}{*}{Personalisation} & \multirow{2}{*}{Data Extraction} &\{name, age, gender, weight, height,\\
    &&location\}&\\
    \hline
    Goal Setting& Data Extraction &\{num\_of\_steps, activity, day\} &\\
    \hline
    \multirow{2}{*}{Reporting}& Data Extraction and & \multirow{2}{*}{\{date, activity, duration, reason\}} & \multirow{2}{*}{\{num\_of\_steps, activity, day\}}\\
    & Contextualisation & &\\
    \hline
    \multirow{2}{*}{Summary} & \multirow{2}{*}{Contextualisation} & & \{date, activity, duration, reason\},\\ 
    &&&\{step\_count, date\}\\
    \hline
    Exercise Coach & Contextualisation & &\{session\_id, exercise\_id, step\}\\
    \hline
    \end{tabular}
\end{table*}

\begin{table*}[!t]
    \caption{Goal Setting Template filling}
    \label{tbl:goal_setting}
    \centering
    \ra{1.2}
    \begin{tabular}{@{\hspace{5pt}} l @{\hspace{15pt}} l @{\hspace{15pt}} l @{\hspace{5pt}}}
    \hline
    \multirow{2}{*}{Agent} & \multirow{2}{*}{User response} & Reporting Template\\
    &&\{num\_of\_steps, activity, day\}\\
    \hline
    What type of goal do you want to set?& I would like a Step Goal& \{?, none, ?\}\\
    $\,\to\,$How many steps do you plan to &8000 steps & \{8000, none, mon\} ... \{8000, none, sun\}\\
    complete a day?&&\\
    \hline
    What type of goal do you want to set?&Activity Goal& \{none, ?, ?\}\\
    $\,\to\,$What are the activities you have & I will be swimming on Monday. & \{none, swimming, mon\}\\
    planned this week?&Then some walking and golf on & \{none, golf, thu\}, \{none, walking, thu\}\\
    &Thursday and Friday&\{none, golf, fri\}, \{none, walking, fri\}\\
    \hline
    \end{tabular}
\end{table*}

\section{Co-creation} \label{sec:co}
Several workshops were organised with intended stakeholders to gather and shape ideas on key functional requirements for a digital conversational intervention.

\subsection{Co-creation Workshops}

\begin{figure}[ht]
\centering
\includegraphics[width=0.45\textwidth]{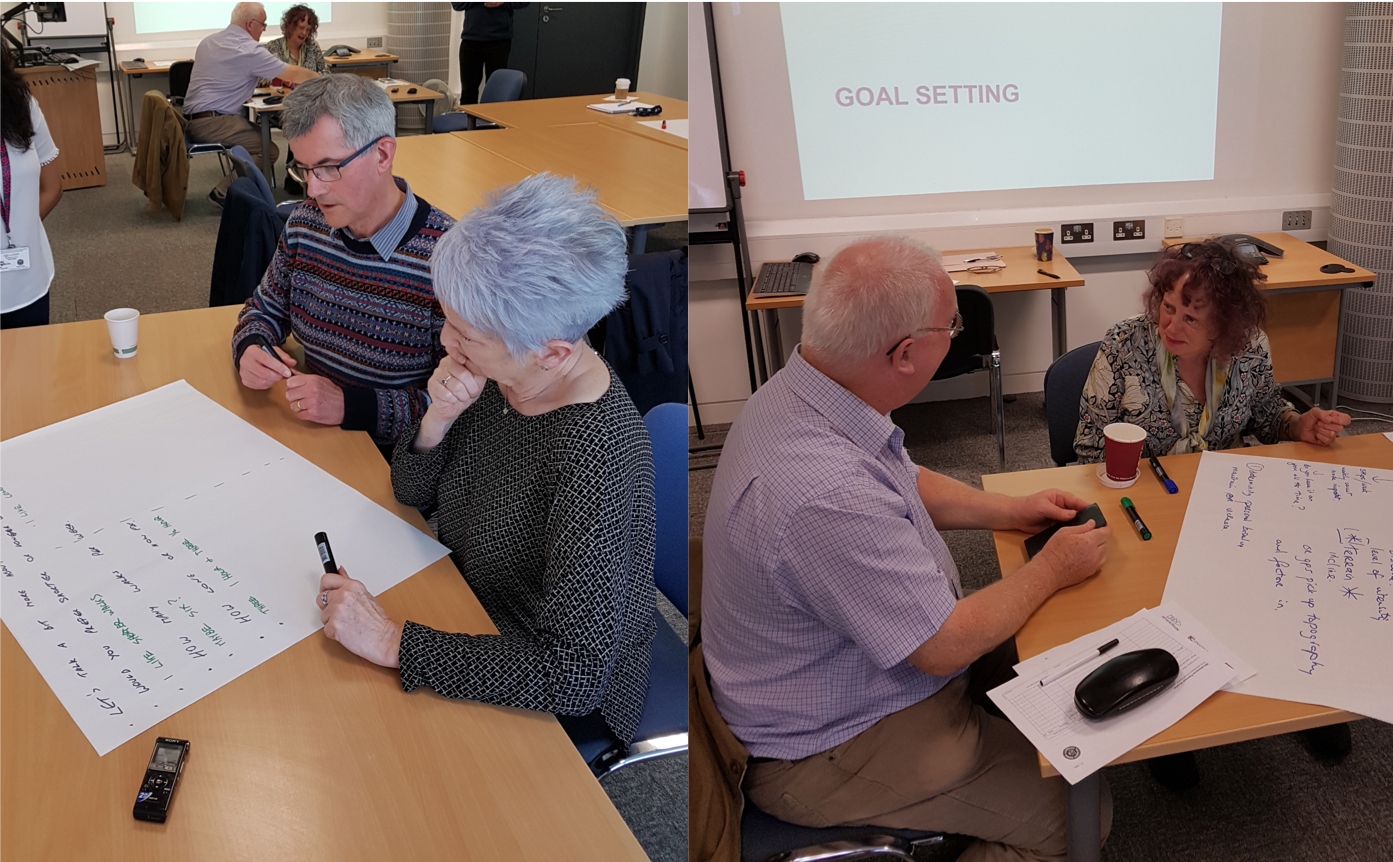}
\caption{Co-creation Workshops}
\label{fig:cocreates}
\end{figure}

The aim of these workshops was to identify and iteratively refine the skills that are expected in a conversational agent that encourages physical activities for older adults~\footnote{
The Robert Gordon University School of Health Sciences Ethics Panel granted ethical approval for the co-creation workshops (SHS/19/04)}.
To allow for an iterative design process, we held 3 co-creation workshops, each one-month apart. Participants (above sixty years) were recruited from various community locations using posters and gatekeeper e-mails.
After reading participant information sheets and discussing the study with a member of the research team, 8 participants (6 male, 2 female) volunteered to take part and attended between 1 and 3 workshops each. 
The workshops used participatory methods (see Figure~\ref{fig:cocreates}) informed by the authors of~\cite{leask2019framework} and iterative refinement methods from~\cite{augusto2018user}. 

Workshop 1 introduced participants to the study and the concept of voice based conversational interventions, and sought their views on proposed features of the intervention (e.g. goal setting, reporting). In Workshop 2, participants formed groups to further explore features which were identified as important in the first workshop. These included the different formats of conversation (i.e. textual, voice) and the design of conversation flows (i.e. the conversational interaction that would allow a user to record their daily activities and different forms of goal setting). Workshop 3 reviewed the features discussed in previous workshops. It also gave participants the opportunity to propose a name for the mobile application. They selected the title ``FitChat'', inspired by the Doric term ``Fit like?'' (Hello, How are you?). 

In order to facilitate co-creation and design of conversational flows we made use of role playing activities amongst workshop participants to understand expectations~\cite{Matthews:2014}.
Specifically we organised participants into pairs, whereby one was asked to play the role of a wizard (or conversational agent) whilst the other plays the human. We were particularly keen to observe the forms of natural dialogue that transpired between each pair. 
We used \emph{intents} identified during co-created activities to provide the needed focus for a given role playing task. 
Whilst these \emph{intents} were co-created during the initial stages of the workshops, they also form the main components of functionality in the final prototype. 

\subsection{Intents}
The goal of an \emph{intent} can be to extract data from the user, present information to the user or a combination of both~(see Table~\ref{tbl:temp}). For instance the \emph{Personalisation} intent gathers information from the user, and  the \emph{Summary} intent provides information. Thereafter the conversational design task is to develop the dialogue to facilitate this flow of information using a template.

For each \emph{intent} that requires information to be gathered (from the user), we define a conversational 'template'. 
This template guides a single conversational interaction between the user and the conversational agent (i.e. the data to be extracted).
Slots in the template are populated by reasoning with the conversational content. This requires information extraction and natural language understanding heuristics to recognise and extract entities from conversation. 
See Table~\ref{tbl:goal_setting} for an example which uses the Goal Setting intent to gather information about the type of goal. 
In essence conversational interactions progress with the aim of template slot filling. 
In Table~\ref{tbl:goal_setting}, a ``?'' indicates mandatory slots and ``none'' indicates optional slots. 
A single conversation can create multiple instantiations of a template if required. For example, if the user wished to set multiple goals instead of a single goal, then the dialogue needs to enable gathering facts about those multiple goal activities (see row two in Table~\ref{tbl:goal_setting}). 
It is important to note that through template slot filling we are able to improve contextualisation of conversations.

\begin{figure*}[ht]
\centering
\includegraphics[width=0.84\textwidth]{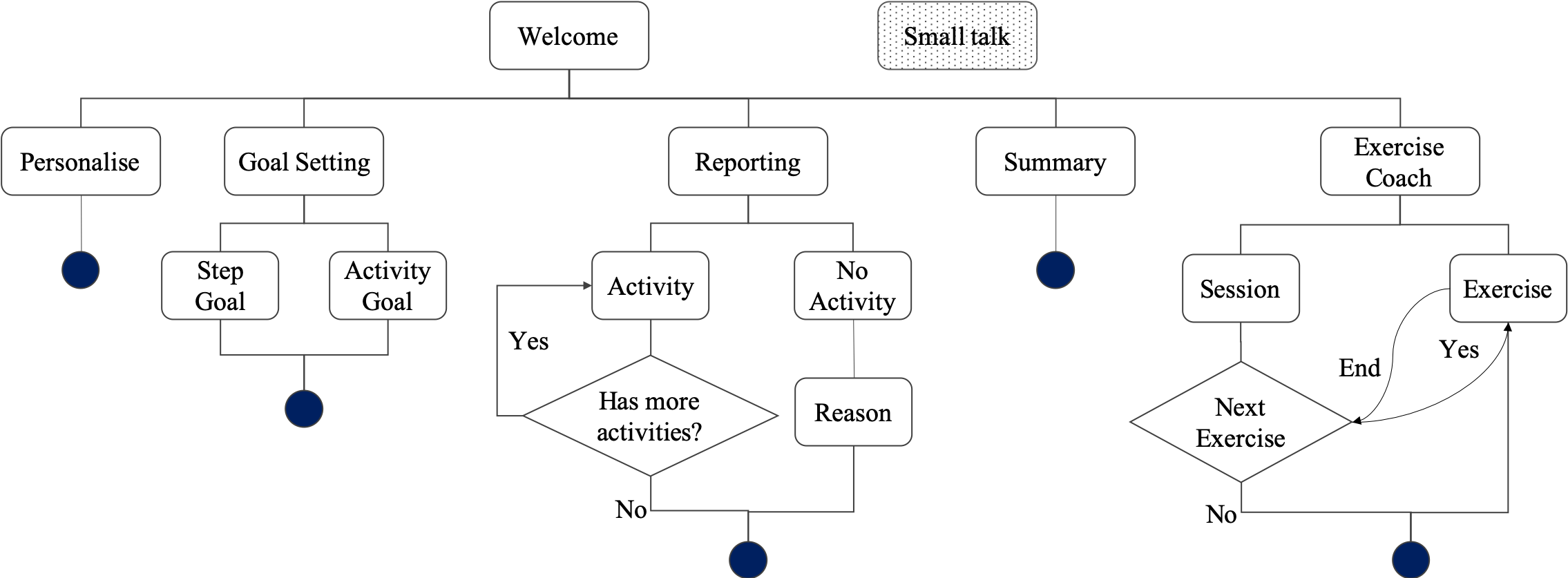}
\caption{Intent Design conversational flow pathways.}
\label{fig:intents_ex}
\end{figure*}
The design of the high-level dialogue structure itself is illustrated in Figure~\ref{fig:intents_ex}. 
Here a \emph{Welcome} intent is used to direct the conversation towards five alternative intents. 
Additionally we also adopted a pre-existing \emph{small talk} (or chit-chat) intent to maintain natural flow of conversation by enabling the user to converse about random topics (if needed).
Note this is not shown in Figure~\ref{fig:intents_ex}.
Together these intents drive core usability functionality of the system, 
whilst application specific intents such as 
Personalisation, Goal Setting, Reporting, Summary and Exercise Coach
are directly related to self-management of physical activity levels. 

\subsubsection{Personalisation}
The goal of the Personalisation intent is to extract personal demographic data from the user to provide a personalised experience throughout the application. As such, it uses a question-answer format to extract information about the user (such as name, age, height, weight, etc) and populate slots in the Personalisation template. This intent is likely to be used only once per user as part of the setup process.

\subsubsection{Goal Setting}
The Goal Setting intent extracts a user's physical activity goals for the upcoming week.
Following from behaviour change theory~\cite{michie}, the idea is to enable users to be conscious (e.g. voicing it) about the specific goals being set as a form of commitment to positive behaviour change.
During co-creation workshop 2, through a group activity, users identified the limitations of fitness apps in recognising physical activities that go beyond ambulatory activities (for instance activities like dancing or golf). 
Accordingly they proposed two types of goals; steps goal and activity goals. The aim of the conversation then is to facilitate the user to set goals of both types in order to account for all types of activities during the week.

The conversational agent starts the conversation by understanding the type of goal the user wants to set, then guides the user towards providing information required by the template. Doing so requires representations that can support true logical forms (LFs) that employ operators (e.g., and, or, equals, if-then-else, etc.) rather than only a flat attribute and value representation.
For a step goal, the template requires the number of steps the user plans to complete each day of the week. For an activity goal, the template requires one or more activities and the respective day for each activity (see Table~\ref{tbl:goal_setting}). 
Further the dependencies between the specific activities and days of the week may shape conversation flows in other intents (such as the Reporting on completed activities discussed next). 

\subsubsection{Reporting}
The Reporting intent is aimed at enable self-reporting of activities for the purpose of self management. 
Depending on the type of goal being set the conversational agent must initiate a contextually relevant dialogue with the intent of extracting activities the user had undertaken during the day.
For this purpose the data obtained from the Goal Setting template 
is retrieved to form the context of the conversation. 
For instance if the user had indicated that she was playing golf on Thursday's then the Agent will be able to ask how she got on with that activity on a Thursday.

In Figure~\ref{fig:intents_ex} we can see that there are two main conversation pathways a user can be directed towards: either the user has performed one or more physical activities and they record them with the Agent, or the user has not performed any physical activities and records a reason (e.g. a barrier). 
At the end of a conversation pathway, the Agent is designed to respond with an appropriate motivational message. 
These are selected based on the pathway and the reason (i.e. barrier) for when a user has not performed an activity.
For instance an example messages such as "Regular physical activity is really good for your well being. Try and fit activity into times of the day that are most convenient to you." addresses a barrier such as "I had no time this week", which is relevant to a user with an activity goal involving "steps". 
An example that is more suited to a bad weather related barrier with goals other than steps would be "Develop some activities that you can always do regardless of the weather, e.g. dancing, stair-climbing or an exercise DVD. Have a plan B for bad weather”.

\subsubsection{Summary}
The Summary intent is designed to provide the user a report of last week's goal achievement, hence it is a simple query-response task. Many participants of the co-creation workshops expressed that listening back to a record of what physical activities they performed is rewarding. Accordingly we use the data gathered by the Goal Setting and Reporting templates to contextualise this conversation by highlighting goals achieved and reported physical activities. A motivational message is added at the end of the summary to encourage the user to maintain or improve their performance next week. 

\subsubsection{Exercise Coach}
\begin{figure}[ht]
\centering
\includegraphics[width=0.40\textwidth]{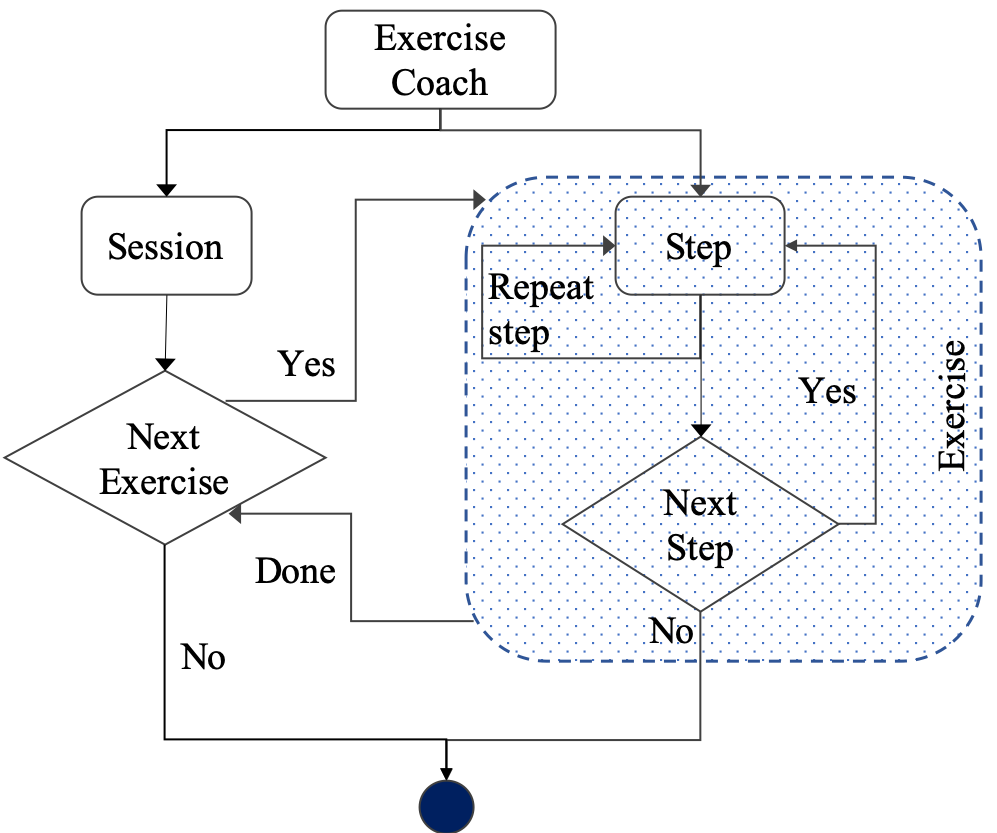}
\caption{Exercise Coach conversational flow pathways.}
\label{fig:ex}
\end{figure}
The purpose of the Exercise Coach intent is to guide users to perform exercises by providing exercise steps through read-aloud instructions in a conversational format. 
The Exercise Coach intent currently supports three exercise sessions: balance, flexibility and strength. 
Exercises are organised in two alternatives formats:
a single exercise at a time; or a set of exercises curated by a physiotherapist as a single session to be performed. 
Each session contains four to six exercises related to that category curated by a physiotherapist. 

A detailed view of the conversation flow is illustrated in Figure~\ref{fig:ex}.
The conversation design also allows these exercises to be performed by the user individually.
Parts of this intent can be viewed as an instructional read-aloud function,  
with the added functionality to enable voice-commands to enable the user to interact in real-time. 
Typical voice commands include: 
\emph{next} (move to next step);  \emph{repeat} (repeat the current step); and \emph{all steps} (read out the entire exercise).

\subsection{Motivational Message Bank}
We adapt a similar methods to corpus-based methods used in literature~\cite{morris2018towards} to develop a Motivational Message bank 
organised under two main categories. Firstly we create a bank of general motivational messages for when the user reports on completed activities; and secondly a set of messages to be used when a user does not perform a planned activity due to a specific barrier. Messages in the barriers category are grouped under six barriers that are commonly found in literature (these include Family, Support, Tiredness, Work, Time and Weather). 
The aim is to deliver a personalised and empathetic response when a user is unable to perform an activity due to a specific barrier. This message bank is integrated with the Reporting and Summary intents.

\section{Think Aloud Sessions}
\label{sec:think}
Think Aloud sessions were planned to evaluate the first prototype of FitChat. The outcomes of these sessions were thematically analysed and presented here. 

\subsection{Study Design}
On completion of the co-design workshops, the app was further refined and participants invited to take part in “think aloud” sessions in order to provide real-time feedback on the intervention. 
Think aloud methods are frequently used for usability testing of e-Health applications~\cite{maramba2019methods}, and involve participants literally thinking aloud whilst they perform a task, or immediately afterwards~\cite{eccles2017think}.
Seven participants took part in five think aloud sessions (two sessions were conducted with two participants in each). Only four participants had previously participated in co-creation workshops. 
During the sessions, participants adhered to the following protocol to explore the features of the application with minimal input from the researchers:
\begin{enumerate}
    \item Login using google sign-in
    \item Set a step goal with the FitChat bot
    \item Explore Traxivity (day and week views) and set a step goal manually
    \item Report an unplanned activity and get a summary of last weeks activities 
    \item Set an activity goal, report a planned activity and get a summary of last weeks activities 
    \item Report that the user did not perform a planned activity
\end{enumerate}

\noindent The think aloud sessions were audio recorded and data was thematically analysed by two researchers and arranged in to six themes.

\subsubsection{Goal Setting Feature:}
\begin{figure}[!t]
\centering
\includegraphics[width=0.49\textwidth]{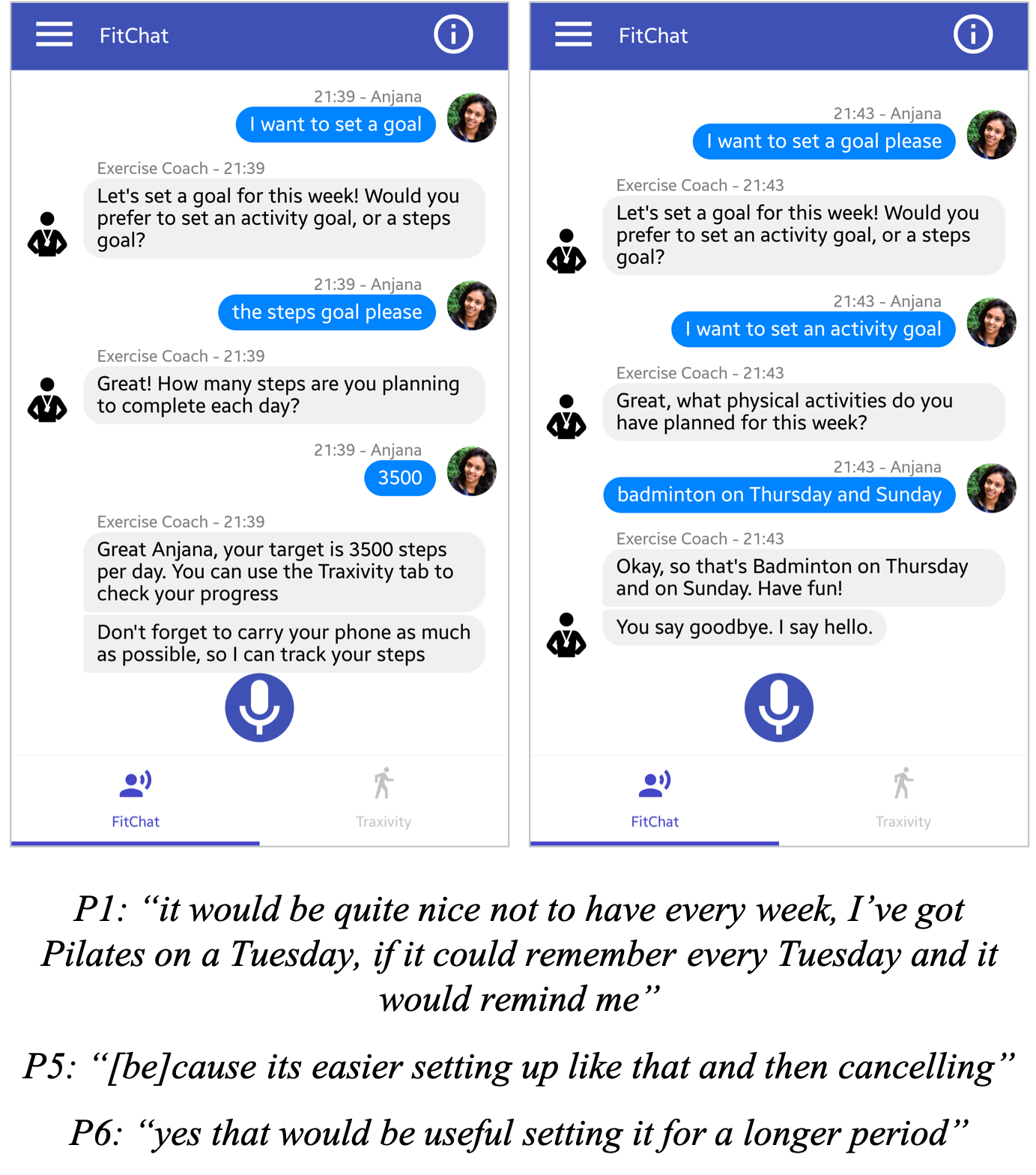}
\caption{Goal Setting}
\label{fig:gs}
\end{figure}

Participants generally responded positively to the goal setting feature. However, they commonly said ``I want to set an activity/steps goal'', for example, rather than ``I want to set a goal'' as instructed in the step-by-step dialogue. Participants commonly suggested that the goal setting feature could be improved if their goals could be stored for longer than one-week, as we illustrate in Figure~\ref{fig:gs}.

\subsubsection{Traxivity Feature:}
\begin{figure}[!t]
\centering
\includegraphics[width=0.49\textwidth]{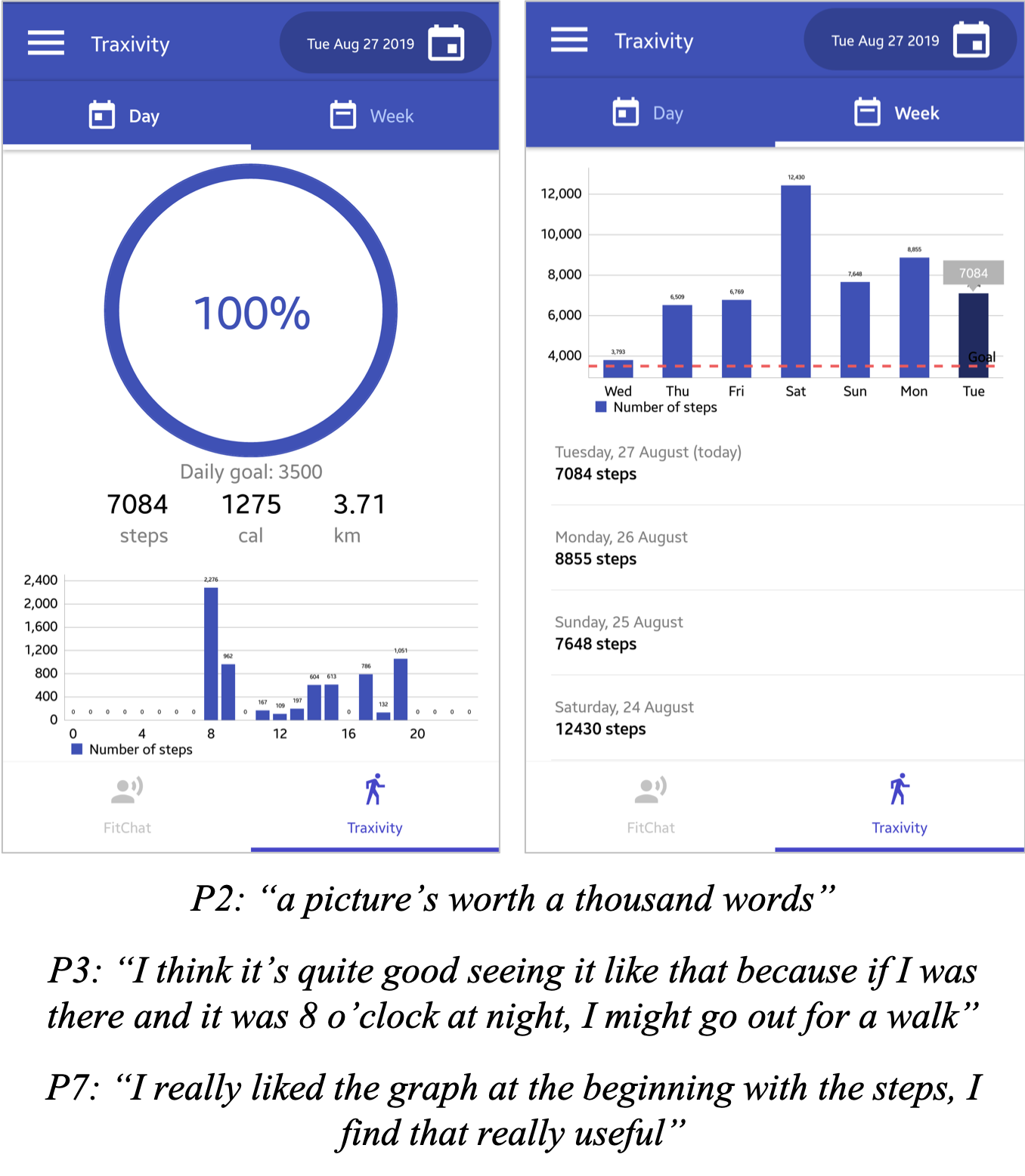}
\caption{Traxivity}
\label{fig:tx}
\end{figure}

Participants commonly liked the visualisation of their steps through the graphs and charts provided by Traxivity (Figure~\ref{fig:tx}).

Some participants suggested that this feature would be motivational with regard to engaging in physical activity:

\begin{quotation}
P1: \textit{``that’s very motivating''}
\end{quotation}
\begin{quotation}
P7: \textit{``I like that, yeah, [be]cause that would maybe make you do something in the evening''}
\end{quotation}

\noindent In terms of improvements to this feature, visualisation of non-stepping activities was commonly proposed, e.g.

\begin{quotation}
P6: \textit{``even just a graph to show how often you were active''}
\end{quotation}

\subsubsection{Reporting Feature}
\begin{figure}[!t]
\centering
\includegraphics[width=0.49\textwidth]{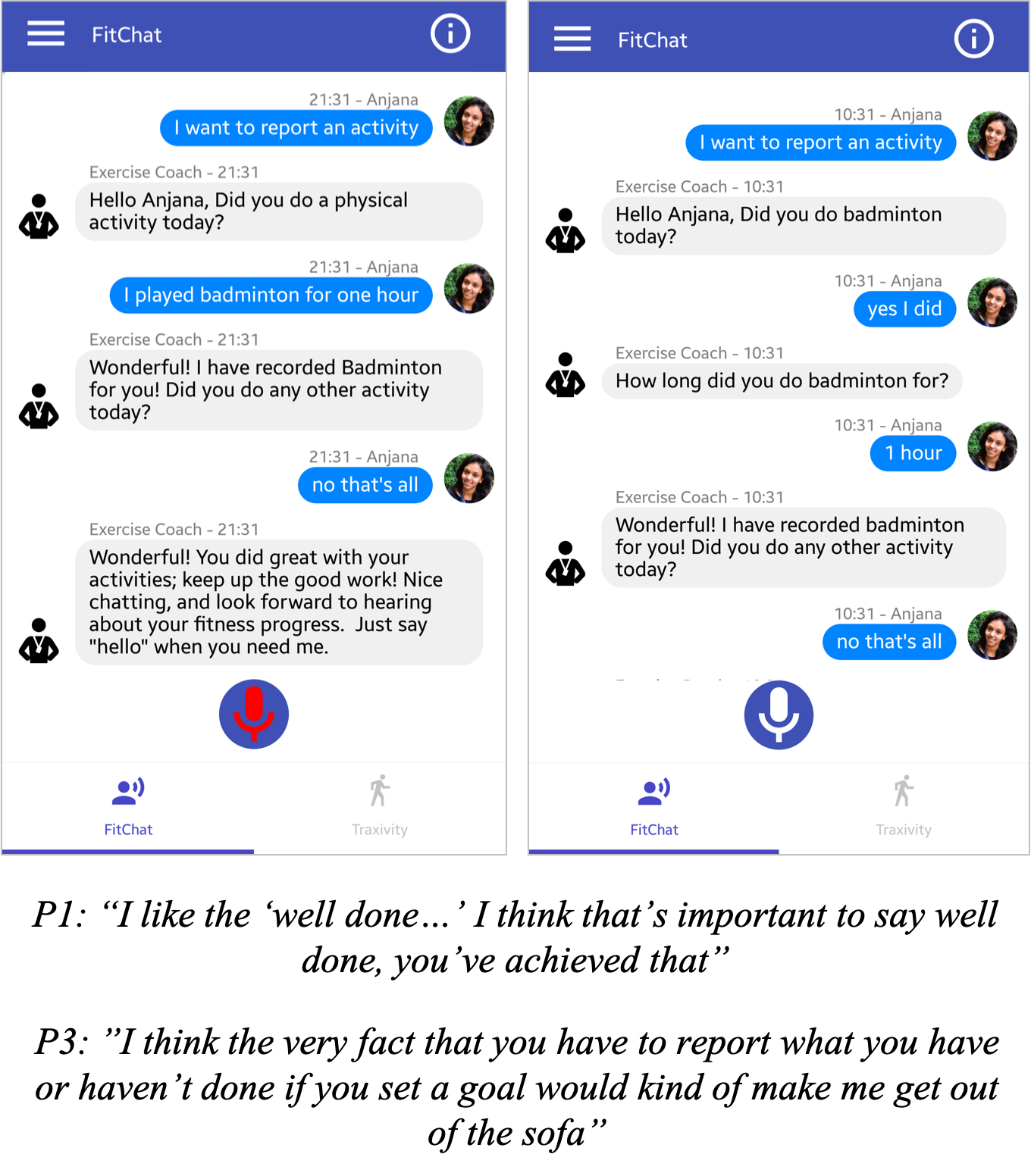}
\caption{Reporting}
\label{fig:report}
\end{figure}

Participants responded well to the reporting feature and were particularly impressed that the app could link their reported activities with their activity goals for the day. Some expressed generally positive responses about the reporting feature (see Figure~\ref{fig:report}).

Most participants suggested that further detail should be added to the activities that are reported to the FitChat application. This commonly included the conversion of reported activities to a unit of measurement e.g. calories burned. Further to this, several participants outlined that they would like to differentiate between the intensity of activities.

\begin{quotation}
P1: \textit{``slightly more detail, in that was it a leisurely swim, I don’t know how you’re going to phrase it, or was it a power swim?''}
\end{quotation}
\begin{quotation}
P2: \textit{``I’m not saying I dislike it but not knowing how much an activity counts for''}
\end{quotation}
\begin{quotation}
P3: \textit{``we could walk 10,000 steps but strolling does nothing for us, so does that come into it?''}
\end{quotation}

\begin{figure}[!t]
\centering
\includegraphics[width=0.5\textwidth]{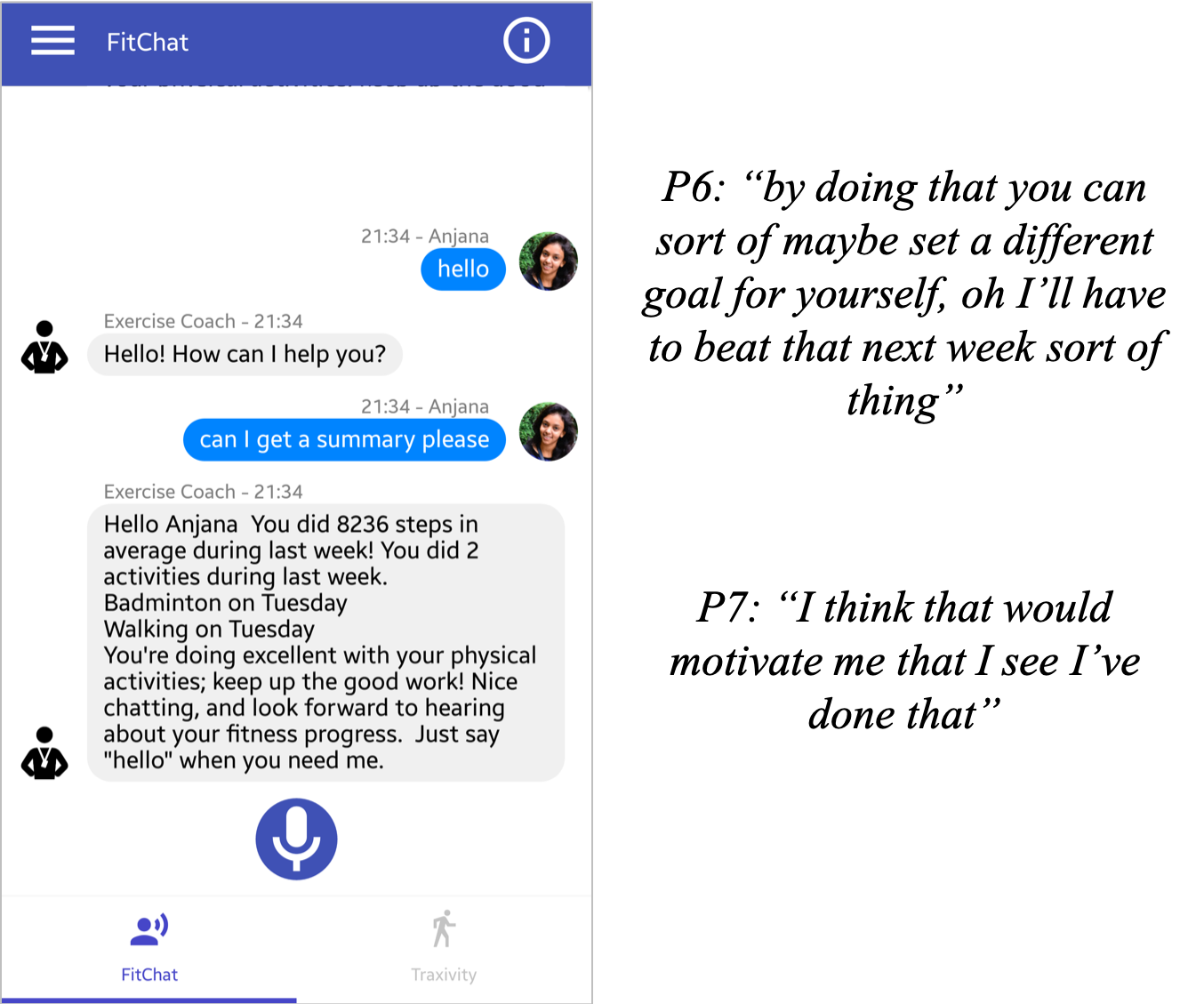}
\caption{Summary}
\label{fig:summary}
\end{figure}

\subsubsection{Summary Feature:}
Feedback was overwhelmingly positive for this feature, with several participants outlining its motivational aspect (see Figure~\ref{fig:summary}). It was also suggested that it would be useful if this feature included a comparison of summaries in order for participants to reflect on differences in physical activity, for example:
\begin{quotation}
P4: \textit{``it might say well done last week you did 40 minutes, the week before you did whatever''}
\end{quotation}

\subsubsection{General Feedback/Impressions:}
Identification of effective skills to minimise complexity is important usability aspect for conversational bots. A complex solution will introduce a learning overhead to the user which is not desirable specially among the older adults. At times, participants did not intuitively converse with the expected phrases/terms required to interact with the application; 
however, they quickly learned the terminology during the think aloud sessions.   
They acknowledged that the app was easy to use once they were familiar with the terminology but expressed that a more complex system would discourage them:

\begin{quotation}
P1: \textit{``you would get into this lingo because it’s obviously got lingo that you have to tap into''}
\end{quotation}
\begin{quotation}
P3: \textit{``I think we would learn very quickly''}
\end{quotation}
\begin{quotation}
P4: \textit{``it takes a wee while to know the tricks like''}
\end{quotation}
\begin{quotation}
P6: \textit{``its easy, simple to use, it’s just because at the moment its very word specific and restrictive'' }
\end{quotation}

\noindent Several participants suggested that it would be useful to include a form of guidance to support the user with the app terminology:

\begin{quotation}
P2: \textit{``people might be more likely to look at a short YouTube clip, how it works, what you can do, how it helps you''}
\end{quotation}
\begin{quotation}
P3: \textit{``maybe we should be given a few key words like ‘record and ‘report’, words that it likes''}
\end{quotation}
\begin{quotation}
P4: \textit{``I think if you maybe did a short introduction just to give some tips''}
\end{quotation}

\noindent With regard to the conversational component of this app, the feedback was largely positive:
\begin{quotation}
P1: \textit{``I think it’s one stage up from a Fitbit definitely because you can interact with it''}
\end{quotation}
\begin{quotation}
P3: \textit{``it’s quite powerful the speaking bit''}
\end{quotation}
\begin{quotation}
P4: \textit{``because you have to speak and listen , aye, you’re almost admitted to yourself, it’s a bit more, you take it more to heart than just clicking a button``} 
\end{quotation}
\begin{quotation}
P6: \textit{``talking is a lot simpler I think, certainly with older people''}
\end{quotation}
\begin{quotation}
P7: \textit{``conversation is far more motivating''}
\end{quotation}

\noindent Some participants suggested that improvements to the conversation could include a greater variety of responses. Some also expressed that they would prefer more informal language:
\begin{quotation}
P4: \textit{``[be]cause you’ll pay attention and kind of look forward to it’s going to be a different form of praise every week''}
\end{quotation}
\begin{quotation}
P5: \textit{``the only criticism I have, is just a small criticism, eh when I first read my report, that feedback, I got all these fancy words''}
\end{quotation}

\noindent In terms of future app usage, the feedback was again positive. Participants were keen to use the app and several stated that they would recommend it to others. Most participants thought that they would use this app long-term; however, a few noted that they would need to try it first. 

\noindent Only one participant demonstrated reticence with regard to using a conversational app:
\begin{quotation}
P5: \textit{``has anybody commented that our generation\dots are not easy to speak to a machine''}
\end{quotation}
\begin{quotation}
P5: \textit{``I just feel\dots no embarrassed but uncomfortable speaking to a phone when there’s people going about''}
\end{quotation}

\subsubsection{Recommendations for Additional Features}
Most participants recommended notifications on the app to remind them of their activity goals, for example:
\begin{quotation}
P2: \textit{``if there was a notification from that, you’ve only done 20\% today’ and its 4 o’clock…… that would motivate me if it were in there''}
\end{quotation}
\begin{quotation}
P4: \textit{``I was wondering about some kind of reminder''}
\end{quotation}

\noindent Further suggestions included rewards (e.g. certificates), sharing achievements with other users or through social media, suggestions of activities in local areas, a playback feature and recommendations for steps goals.

In summary, the themes emerged in the above analysis highlights that features Goal setting, Traxivity, Reporting, and Summary are highly accepted by the users. Accordingly we suggest that they are identified as the most effective conversational skills in a fitness chatbot. 
Features such as goal setting, reporting and summary receiving mostly positive feedback from the users suggests that the voice based activity scheduling and reporting has a positive effect on encouraging physical activities. Many users expressed that ``saying out loud'' the planned activities will encourage them to follow-up on their plan. In addition, ``saying out loud'' the activities they performed and ``listening back'' to the activities they performed within the week is motivating and powerful compared to text or choice based input/output formats. 
With regards to the engaging nature of conversation, the users had mixed reactions. Users suggested improving the message banks to avoid repetitiveness during long term use and they were aware that they are conversing with a bot.

With these results we identify many paths of improvements for FitChat, but most importantly we recognise that the voice based conversational bots are largely accepted by the end-users towards encouraging physical activities. Despite the novelty of voice based conversational AI, user responses highlighted the enthusiasm for learning new technology and understanding the iterative development required to co-produce a solution that is acceptable to all stakeholders. We believe these insights are invaluable for the next phase of FitChat. 

\section{Related Work}
\label{sec:related}

Conversational agents have been tried as intervention delivery methods in many healthcare application domains including mental health~\cite{morris2018towards,inkster2018empathy,suganuma2018embodied}, Asthma~\cite{rhee2014mobile}, 
weight loss and obesity~\cite{stein2017fully,addo2013toward}, 
physical activity and diet~\cite{fadhil2019assistive,fadhil2017adaptive}, medication adherence~\cite{fadhil2018conversational} and alcoholism~\cite{lisetti2011toward,lisetti2013can}. 
Many of these use smart phone applications where the user can only respond either by selecting an option from a number of choices or through free text entry. 
Alternatively early research explored the use of a web based avatars to integrate voice and emotions into intervention delivery~\cite{lisetti2011toward,lisetti2013can}. 
However voice based conversational interfaces in the form of chat-bots are considered to be more natural and intuitive, compared to these traditional web based avatars. 
However, a comprehensive vocabulary is essential to ensure that the learning curve is manageable without requiring the user to memorise key phrases 
to carry on a dialogue with the tool. 

A recent evaluation of Wysa~\cite{inkster2018empathy}, a text/multiple-choice empathetic AI chat-bot for mental well-being, focused on analysing user acceptance of conversational agents. 
Their findings suggests that a majority of 67\% found Wysa to be a ``Favourable Experience'' compared to 32\% who found it to be a ``Less Favourable Experience''. 
Additionally users preferred to respond by clicking on options given by the app when compared to entering free text.

Recent literature suggests a corpus-based approach for enforcing empathy into text/choice based conversational bots~\cite{morris2018towards}. 
A corpus is curated with empathetic responses that will be used by the conversational agent when responding to a user. They measure the acceptability of empathetic responses presented by the bot compared to responses presented by a peer and finds that users accept bot responses 79\% of the time. We find this as an interesting approach, to build a response bank in order to motivate the users to improve their physical activity levels.

Lark~\footnote{https://www.lark.com/outcomes} is a well-known text/choice based Conversational Agent specialised in diabetes management and~\cite{stein2017fully} evaluates Lark for user acceptability and satisfaction where users rated the app at 7.9~(average) on a 0-10 scale. These studies suggest that in general conversational agents are widely accepted by the users, but they are limited to text or choice based responses.

In the context of encouraging physical activity with older adults, we argue that the voice is more effective in delivering motivational content. It is noteworthy that older adults are in general not accustomed to free text entry with smart phones. In addition, the recent popularity of home hubs presents an opportunity to build voice based conversational agents for both smart-phone and home-hub platforms simultaneously. Accordingly we plan to evaluate the acceptability of voice as the user and conversational agent response format. We overcome the inaccessibility of existing voice based methods by implementing a smart phone app. In addition our initial feasibility studies suggested that content delivered through text cannot be directly used in a voice platform; alterations are required to adjust the content such that the conversation flow preserves informality and delivers it using the right tone of voice. We have explored these challenges that are less studied in literature through our FitChat application. 

\section{Conclusion}
\label{sec:conc}
In conclusion, we have identified that conversation has great potential to deliver effective Digital Behaviour Change Interventions (DBCIs) to encourage physical activity in older adults. 
To measure this claim, in this work we exploited the advances of conversational AI to build the voice based Conversational bot ``FitChat''. We explored the essential features of a DBCI with older adults from the community through co-creation workshops and evaluated the first prototype through think aloud sessions. Thematic analysis of the think aloud session outcomes suggests that voice is a powerful mode of delivering DBCI which may increase adherence to physical activity regimes and provide motivation for trying new activities. In future we plan to conduct a feasibility study that will evaluate the long term effects of voice based conversation in encouraging physical activities with older adults. 

\section{Acknowledgement}
This project is funded by the GetAMoveOn Network+ (funded by the Engineering and Physical Sciences Research Council, UK (EPSRC) under the grant number: EP/N027299/1) and the SelfBACK Project (funded by the European Union's H2020 research and innovation programme under grant agreement No. 689043).
We would like to thank Valery Burnett and Jess McGowan from Robert Gordon University, UK and Benjamin Picard and Leane Seguin from University of Clermont Auvergne, France who contributed to this project in multiple capacities. 

\bibliography{references}
\bibliographystyle{aaai}

\end{document}